\documentstyle [epsf]{laa}

\newcommand{\eqb}{\begin{eqnarray}}
\newcommand{\eqe}{\end{eqnarray}}
\newcommand{\gesim}{\,\raisebox{-0.4ex}{$\stackrel{>}{\scriptstyle\sim}$}\,}

\newcommand{\nelec}{n_{\rm e}}
\newcommand{\nprot}{n_{\rm p}}

\newcommand{\tsd}{t_{\rm Sd}}
\newcommand{\ush}{u_{\rm sh}}
\newcommand{\ushfr}{u_{{\rm sh},4}}
\newcommand{\eel}{E_{\rm e}}
\newcommand{\eep}{E_{\rm p}}

\newcommand{\gmax}{ E_{\rm e,max}}
\newcommand{\gacc}{ E_{\rm acc}}
\newcommand{\gcr}{E_{\rm loss}}

\newcommand{\cesn}{{\cal E}_{\rm SN}}
\newcommand{\cesnfi}{{\cal E}_{{\rm SN},51}}
\newcommand{\ceel}{{\cal E}_{\rm el}}
\newcommand{\cepr}{{\cal E}_{\rm pr}}
\newcommand{\rc}{\rm r_c}
\newcommand{\tyr}{\rm t_{yr}}

\newcommand{\cLe}{{\cal L}^{\rm e}}
\newcommand{\cQe}{{\cal Q}^{\rm e}}

\begin{document}
\thesaurus{02.01.1; 02.18.2; 02.18.3; 02.19.1; 08.19.5 SN 1006; 09.03.2}
\title{On the high energy non-thermal emission from shell-type supernova remnants}
\author{A. \ Mastichiadis}
\institute{\it MPI Kernphysik, Postfach 10 39 80, D-69029 Heidelberg, Germany}
\offprints{A.~Mastichiadis}
\date{Received 19 October 1995; accepted 28 November 1995}
\maketitle
\begin{abstract}
Shock waves associated with shell type supernova remnants are
considered to be possible sites of cosmic ray acceleration.
Since shocks are capable of accelerating electrons
in addition to protons one anticipates both species to contribute
to the high energy radiation expected from these objects.
Adopting a simple model for particle acceleration
we calculate in a self-consistent manner the time-dependent synchrotron
and inverse Compton radiation of high energy electrons
assumed either to be accelerated directly by the shock wave or to be injected
at high energies as secondaries from the 
hadronic collisions of relativistic protons with the circumstellar material.
We deduce that for standard supernova parameters
the TeV flux produced from neutral pion decay 
is about the same order as the flux expected from directly accelerated
electrons.
\end{abstract}

\section{Introduction}
Shell-type supernova remnants (SNRs) are long thought to be the sources
of the nuclear component of cosmic rays 
up to energies close to the knee 
(see, for example, Axford [\cite{axford81}] and Lagage \& Cesarsky 
[\cite{lagagecesarsky83}]). 
SNRs appear to be one of the few galactic objects which
release enough energy to satisfy the observed
flux of cosmic rays. Also recent developments in the theory of
diffusive acceleration in shock waves provide the required 
theoretical background for such a picture (see the review
by Jones \& Ellison [\cite{jonesellison91}]). 

In addition to protons, electrons are also expected to be accelerated directly
in these shock waves. 
However, as protons are
expected to carry practically all the energy available for acceleration
(usually taken to be 10\% of the total supernova energy)
the energy content in accelerated electrons is not known.
Nevertheless, if one assumes that the measured ratio of cosmic ray electron
energy density to that of protons is the same as the one which is produced 
by the SNRs then we should expect
that about 1\% of the SNR energy available for acceleration goes to electrons.

Despite the theoretical arguments, there is not as yet any
firm observational evidence to support the above picture of cosmic
ray origin. It was suggested that 
a direct proof for the theory of proton acceleration would be
the detection of SNRs at TeV energies 
(Dorfi [\cite{dorfi91}] and Drury et al. [\cite{druryetal94}]--henceforth DAV94).
Relativistic protons undergo hadronic collisions with the ambient matter
producing TeV $\gamma-$rays from neutral pion decay. 
This flux should be, in principle, detectable with present-day Cerenkov 
detectors.

On the other hand, 
there is strong observational evidence that SN 1006 
(Becker et al. [\cite{beckeretal80}],
Koyama et al. [\cite{koyamaetal95}]) shows a non-thermal X--ray spectrum and
the possibility that the accelerated (primary)
electrons can radiate synchrotron X--rays has already been discussed
by Reynolds \& Chevalier (\cite{reynoldschevalier81})
and Ammosov et al. (\cite{ammosovetal94}) --however see Hamilton et al.
(\cite{hamiltonetal86}) for
an alternative explanation. It has been suggested that if the X--rays 
observed from SN 1006 prove to be non-thermal in 
origin, this can only mean that electrons are accelerated
to energies of 100 TeV or more, providing thus for the first time
evidence for very high energy particle acceleration in SNRs.

The above statements, i.e. that TeV radiation is proof of proton
acceleration while the non-thermal X--rays are an indication for
electron acceleration, ignore the fact that each of the accelerated
species can give independently a contribution to the X--ray and
$\gamma-$ray flux. Thus directly accelerated electrons will not only
produce synchrotron X--rays but also $\gamma-$rays from inverse
Compton scattering on any ambient photon field present.
Similarly, accelerated protons will not only produce neutral pions
but also charged pions which, upon decay, will create secondary
electrons. These electrons will radiate just as any directly accelerated 
electron contributing to the total SNR X--ray and $\gamma-$ray flux. 
The question which arises therefore is the following: If SNRs 
turn out to be sources of TeV radiation will the radiation come mostly from
protons ($\pi^0$ decay) or from electrons (inverse Compton scattering)?
And if there is, after all, a non-thermal component in the X--ray regime
will it be mostly due to directly accelerated electrons or due to secondary
electrons produced in hadronic collisions? The above questions are
obviously of some importance to acceleration theories. 

The aim of the present {\sl Letter} is to try to address the questions
posed above. In doing so we
calculate, within the standard SNR particle 
acceleration framework,
the time-dependent
X--ray and $\gamma-$ray fluxes produced by each of the accelerated 
species.
In \S2 we describe the model
we base our calculations on, in \S 3 we give results and
we conclude in \S 4.

\section {The Model}
We base our SNR model on a general picture 
including diffusive acceleration 
(for details of the underlying physics see Dorfi [\cite{dorfi93}] and
Kirk et al. [\cite{kirketal94}]) 
treating both electons and protons as test particles.
We assume that the SNR shock front is moving through the  
interstellar medium with a velocity
$\ush\simeq$ const. in the free expansion phase and
$\ush\propto t^{-3/5}$ in the Sedov phase. 
The Sedov phase starts approximately at a time $\tsd$ when the shock has reached
a radius such that the swept up mass equals the ejected mass, thus 
$\tsd\simeq 210 \cesnfi^{1/3}\rho_{1}^{-1/3}\ushfr^{-5/3}~~{\rm yr}$.
Here $\cesnfi$ is the
total energy released in the supernova explosion in units of $10^{51}$ erg,
$\rho_1$ is the external matter density in units of 1 H-atom/cm$^{3}$ and $\ushfr$
is the shock velocity in units of 10$^4$ km/sec.

We take a simple Bohm type of diffusion 
upstream and downstream of the shock and
we assume that protons as well as electrons are accelerated there.
We determine the maximum energy to which the particles have been accelerated 
after some time {\it t} by simply integrating over the acceleration rate
(see, for example,  Webb et al. [\cite{webbetal84}]) which gives
$\gacc\simeq 3 A(\rc)B_{-5}\ushfr^2\tyr~~{\rm TeV}$
for $t<\tsd$ and 
$\gacc\simeq 630 A(\rc)B_{-5}\ushfr^{1/3}
\cesnfi^{1/3}\rho_{1}^{-1/3}~~{\rm TeV}\simeq {\rm const}$
for $t\ge \tsd$.
Here $A(\rc)={(\rc-1)}/{\rc(\rc+1)}$ with 
$\rc$ being the compression ratio of the shock,
while $B_{-5}$ is the magnetic field strength given in units of 
$10^{-5}$~G. Especially for the case of the accelerated electrons
the maximum energy the particles obtain can be limited by losses.
We estimate therefore
the critical energy $\gcr(t)$ at which the acceleration rate
can be balanced by the loss rate (synchrotron and inverse Compton
combined) and set $\gmax(t)={\rm min} [\gacc(t),
\gcr(t)]$ and this is shown in Fig. 1 as a function of the magnetic field $B$.

\begin{figure}[t]
\epsfxsize=7.5 cm
\epsffile{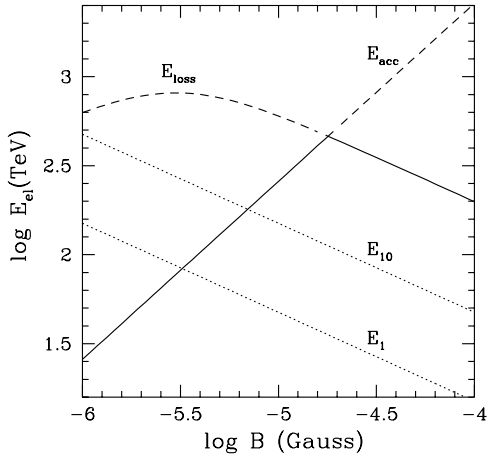}
\vspace{-0.8 cm}
\caption{Maximum energy of electrons as a function of the magnetic
field at shock $B$. $E_{\rm acc}$ is the maximum energy the particles
can attain when losses are neglected -- see text.
$E_{\rm loss}$ gives the critical 
energy
at which the acceleration rate can be balanced by the energy loss rate.
Thus the maximum possible energy is given by the
solid line.
$E_1$ and $E_{10}$ are the energy electrons should have in order to
radiate synchrotron X--rays of energies 1 and 10 keV respectively}
\vspace{-0.3 cm}
\end{figure}

\vspace{-0.1 cm}
\subsection{Emission due to primary accelerated electrons}
We calculate the radiated spectra from the relativistic electrons
once these have escaped downstream of the shock. 
We treat this problem in a self-consistent manner
by solving first a time dependent kinetic equation
for the relativistic electron distribution function $\nelec(\eel,t)$
\eqb
{{\partial\nelec(\eel,t)}\over{\partial t}}=\cQe_{\rm prm}(\eel,t)+\cLe(\eel,t).
\label{ekinetic}
\eqe
Here $\cLe(\eel,t)$ denotes the electron losses which are taken to be due to
synchrotron radiation and inverse Compton scattering, as well as
losses due to adiabatic expansion 
\footnote{In dense media electron bremsstrahlung will become
important, however we restrict our analysis here in shocks propagating
in the ISM where the densities involved allow us to neglect the above process.}
(for the relevant expressions used
see Mastichiadis \& Kirk [{\cite{mastichiadiskirk95}}] and references
therein). 
The source of high energy electrons is given by
the rate $\cQe_{\rm prm}$ of primary electrons that escape downstream. 
For them we assume
a power law spectrum in energy
and write 
$\cQe_{\rm prm}(\eel,t)=
Q_{e,0}(t)\eel^{-s}$ with $\eel < \gmax(t)$ and $s=(\rc+2)/(\rc-1)$.
This approximation is known as the ``Onion-shell-model'' (Bogdan \& V\"olk 1983).
For determining $Q_{e,0}(t)$ we follow Drury ({\cite{drury92}})
and assume that a part of the total energy flux through the shock
goes into accelerating electrons. Thus we write
\eqb
{{4\pi}\over 3}R(t)^3\int d\eel\eel Q_{\rm e,prm}(\eel,t)=4\pi\xi
[R(t)]^2{{1}\over{2}}\rho [u_{sh}(t)]^3,
\label{eflux}
\eqe
where $\rho$ is the ambient matter density and $\xi$ is a 
proportionality constant to be determined.
For steep electron
spectra (i.e. $s>~2$) as the ones we will examine here this energy
flux is  $\propto t^2$ for $t< \tsd$ and $\propto t^{-1}$
for $t>\tsd$. 
From this form of source function it becomes evident that the bulk of
energy output occurs at times $t\simeq \tsd$. 

The solution of the electron kinetic equation (Eqn. {\ref{ekinetic}}) 
gives the distribution function
of primary electrons downstream $\nelec(\eel,t)$ which is an implicit
function of the quantity $\xi$ (Eqn. \ref{eflux}).
We normalize $\nelec(\eel,t)$, and thus determine $\xi$, 
to the fraction of the total supernova energy that goes to accelerated electrons  
$\eta_{\rm el}$ through the equation
\eqb
{{4\pi}\over {3}} R_{\rm SNR}^3\int dE_{\rm e}E_{\rm e}n_{\rm e}
(E_{\rm e},t_{\rm SNR})=\ceel
=\eta_{\rm el}\cesn.
\label{distf}
\eqe
Here $t_{\rm SNR}$ is the time at which the remnant ends its life, 
$R_{\rm SNR}\equiv R(t_{\rm SNR})$, while $\ceel$ is the
energy content in electrons produced in the lifetime of the supernova.
Since most of the power in particles is put into the SNR at times 
$\ll t_{\rm SNR}$, at late epochs
Eqn. ({\ref{distf}}) changes very slowly with time and therefore
it is  insensitive to the exact value of $t_{\rm SNR}$.
However for concreteness we follow Dorfi ({\cite{dorfi93}}) and use 
$t_{\rm SNR}\simeq
1.3~10^6\cesnfi^{11/35}\rho_{1}^{-13/35}$~years.
Once $\nelec(\eel,t)$ is determined we can use it
to calculate both the synchrotron and inverse Compton emissivities
by folding it with the corresponding emissivities.

\begin{figure}[t]
\epsfxsize=7.5 cm
\epsffile{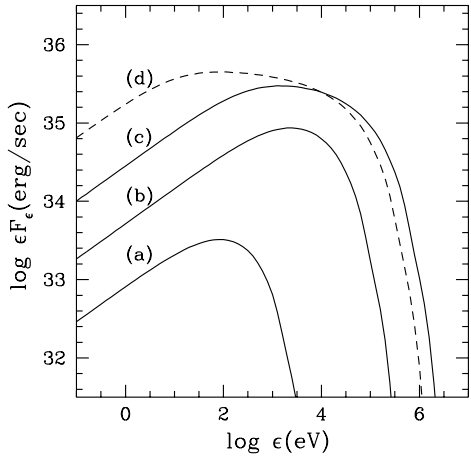}
\vspace{-0.8 cm}
\caption{
Synchrotron spectra produced due to primary accelerated electrons
at $t$=1,000 years after the explosion
as a function of the magnetic field. (a) is for 3 $\mu$G, (b) is
for 10 $\mu$G, (c) is for 30 $\mu$G and (d) is for 100 $\mu$G.
The other parameters are as given in the text. 
}
\vspace{-0.3 cm}
\end{figure}
\begin{figure}[t]
\epsfxsize=7.5 cm
\epsffile{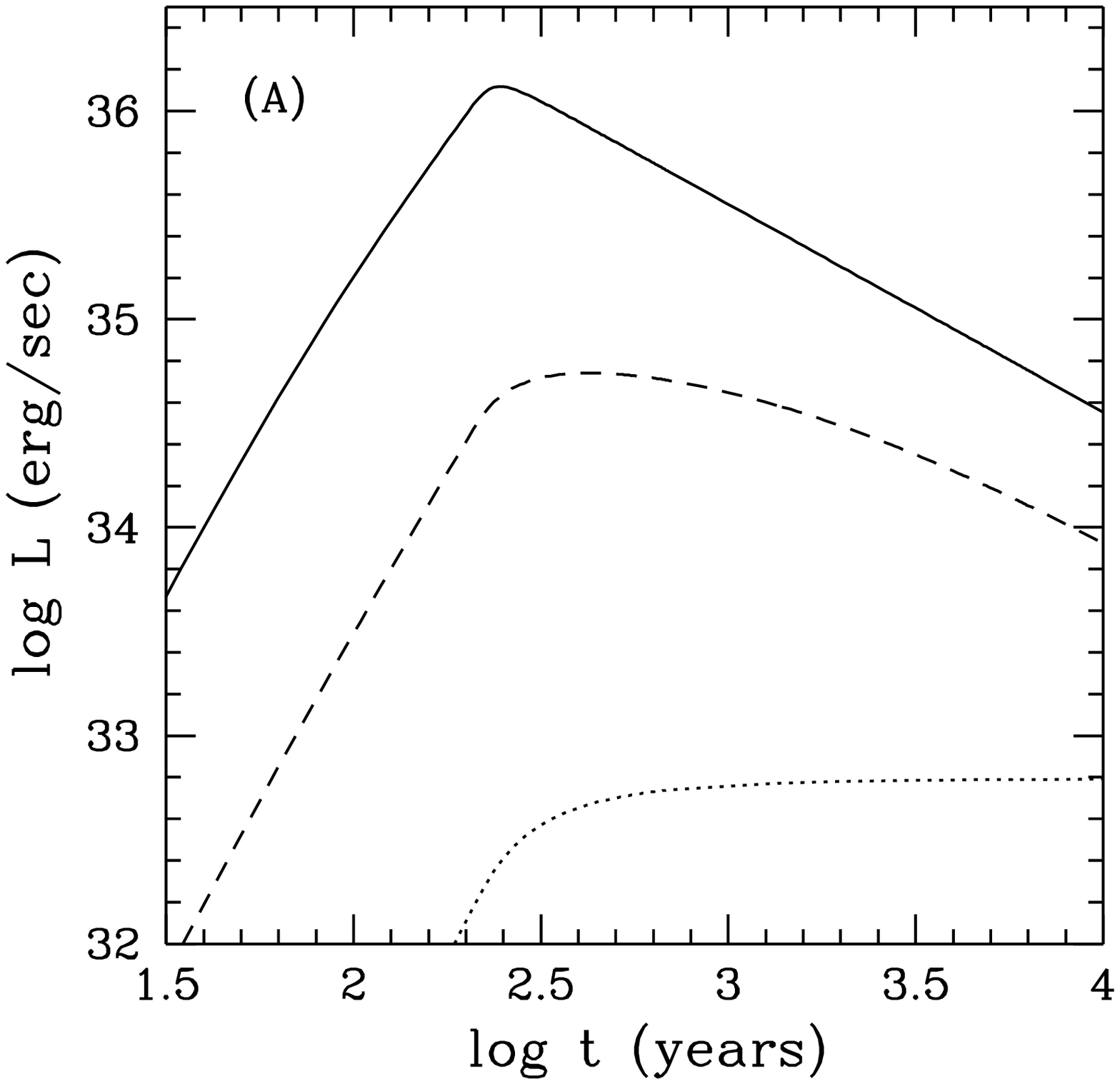}
\vspace{-0.8 cm}
\epsfxsize=7.5 cm
\epsffile{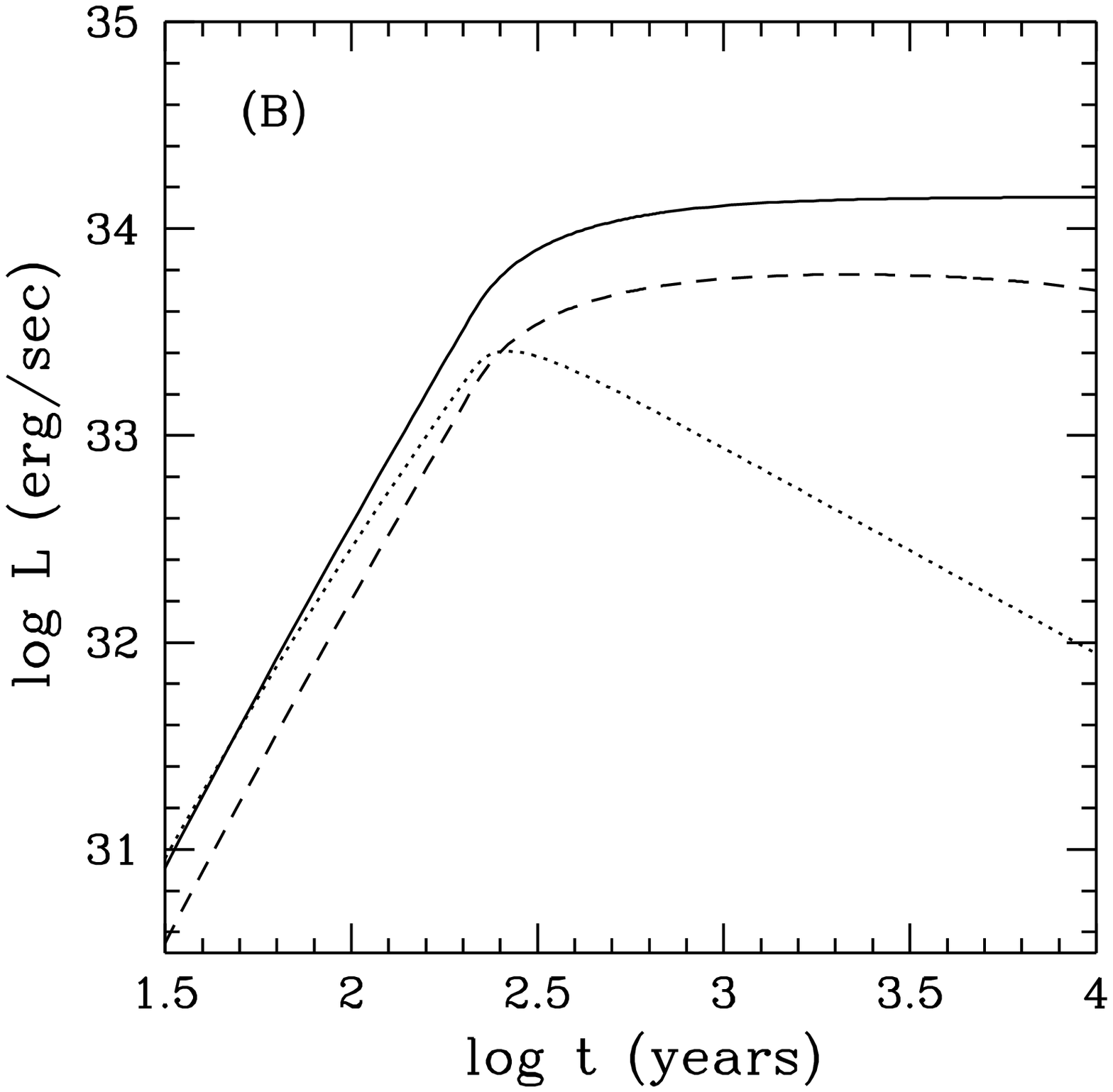}
\vspace{-0.8 cm}
\caption{
{\bf A} SNR synchrotron X--ray lightcurve in the 1-10 keV energy band.
The full and dashed lines are due to radiation of
primary electrons in a magnetic
field of B=100 $\mu$G and 10 $\mu$G respectively,
while the dotted line is due to secondary electrons and B=100 $\mu$G.~~
{\bf B} SNR $\gamma-$ray lightcurve for energies $>$ 1 TeV. The full line is due
to $\pi^0$ decay, while the dashed and dotted lines
are due to inverse Compton scattering of primary
electrons on the microwave background. The magnetic field was taken to
be 3 and 100 $\mu$G respectively.}
\vspace{-0.3 cm}
\end{figure}

\vspace{-0.1 cm}
\subsection{Emission due to accelerated protons}
In the case of proton acceleration we solve a kinetic equation 
similar to Eqn. ({\ref{ekinetic}})
but taking into account only adiabatic losses. 
The source of high energy protons is provided by the proton escape downstream.
In close analogy with the electron case
we introduce a proton
efficiency $\eta_{\rm pr}$ which we relate to the total
energy content in protons $\cepr$ produced in the lifetime of the supernova
by an equation similar to Eqn. (\ref{distf}). The solution of the kinetic equation 
for protons gives their distribution function 
$\nprot(\eep,t)$ which we use subsequently
to calculate the produced spectra of the
$\gamma-$rays  $Q_{\pi\gamma}$ and
secondary electrons $\cQe_{\rm sec}(\eel,t)$ (produced through neutral and charged 
pion decay respectively -- for details on the methods used 
see Mastichiadis \& Kirk [{\cite{mastichiadiskirk95}}]).
$Q_{\pi\gamma}$ is the $\gamma$-ray emissivity considered by
Dorfi ({\cite{dorfi91}}) 
and DAV94. 

For the case of secondary electrons we have
to solve again the electron kinetic equation  ({\ref{ekinetic}})
using $\cQe_{\rm sec}(\eel,t)$ instead of 
$\cQe_{\rm prm}(\eel,t)$
as the source of energetic electrons to obtain
the distribution function of the secondary electrons $n_{\rm e,sec}(\eel,t)$.
This can be used subsequently
to calculate both the synchrotron and inverse Compton emissivities in
a manner similar to the one described in Section 2.1.
Note that once 
$\nprot(\eep,t)$ has been specified, the quantities
$Q_{\pi\gamma}$, $\cQe_{\rm sec}$ and  $n_{\rm e,sec}$
are completely determined.

\section{Results}
We solve numerically the above described three kinetic equations (two for electrons -- 
primary and secondary, and one for protons). We assume 
that the shock 
compression ratio is $\rc=3.73$ which
implies that the accelerated particle index is $s=2.1$. 
This value was chosen because it 
is in agreement with estimates for the possible cosmic ray index at source
-- see DAV94 and Berezhko et al. ({\cite {berezhkoetal94}}). 
Furthermore we assume that 
$\eta_{\rm pr}=.1$  and $\eta_{\rm el}=.001$ while  
$\cesn=10^{51}$ erg, $\ush=10^4$ km/sec and $\rho$=1 H-atom/cm$^3$,
while we leave the magnetic field strength as a free parameter.
Note that our results scale linearly with 
$\eta_{\rm pr}$ and $\eta_{\rm el}$. Also the proton related fluxes 
(i.e. $Q_{\pi\gamma}$ and $\cQe_{\rm sec}$) scale linearly with the
ambient density $\rho$. 

Fig. 2 shows the synchrotron X--ray spectra of primary electrons
at $t$=1,000 years
after the explosion as a function of the magnetic field
which we assume to be uniform in the downstream region.
One notices that both
the flux and the high energy cutoff increase
with increasing magnetic field. The flux increases
because of the dependence of synchrotron emissivity on the magnetic field.
The high energy cutoff increases because one can roughly write
$\epsilon_{\rm max}\propto E_{\rm e,max}^2B\propto B^3$
for those magnetic fields where acceleration is not limited as yet by losses
(see Fig. 1).
From Fig. 2 it is apparent that one
needs to have magnetic fields
larger than 3 $\mu$G for 10 keV X--rays to be produced (see also Fig. 1).
There are some interesting consequences to the spectra for relatively
high magnetic fields ($B\gesim$ 30 $\mu$G). Thus the high energy cutoff 
stops increasing as now $E_{\rm e,max}$ is determined by losses.
Also the photon spectrum shows a break; this reflects a similar
break in the electron distribution function which is
caused by electron synchrotron cooling.
This break is absent for low magnetic
fields as for these fields the synchrotron timescale is long
compared to the age of the SNR.

Fig. 3 shows the expected X--ray and $\gamma$-ray ($>1$ TeV) lightcurves
from a SNR characterised by the parameters given above. 
X--rays are produced by the synchrotron radiation
of electrons while $\gamma$-rays are produced from $\pi^0$ decay
and from electron inverse Compton scattering on the cosmic microwave
background. 
For the assumed values of  
$\eta_{\rm el}$  and $\eta_{\rm pr}$   
we find that 
primary electrons produce higher fluxes than secondaries
both in the X-- and in the $\gamma-$ray regime.
For high magnetic fields ($\gesim$ 50 $\mu$G) 
the synchrotron cooling timescale is short compared to the age of the SNR; 
this means that electrons radiate efficiently
and the synchrotron lightcurve follows essentially the source function
of the electrons (solid line curve in Fig. 3A). When the magnetic
field is lower the above argument is no longer valid 
as the electrons never have the time to cool efficiently
(dashed line curve in Fig. 3A).

In the $\gamma$-ray regime (Fig. 3B)
we find that the inverse Compton contribution from the primary
electrons (dashed line) is of the same order as that of the $\pi^0$ 
$\gamma$-rays (solid line) provided that the magnetic field is not much larger than 10 
$\mu$G. (It is interesting to note that our $\pi^0$ $\gamma$-ray lightcurve
is in excellent agreement with the result of DAV94 despite
the different approaches of the two papers.)
Higher magnetic fields suppress the inverse Compton flux as they put 
most of the radiated power in 
synchrotron.

However we find that TeV observations could tell 
whether the observed spectrum is due to protons or electrons.
In the case of protons one expects a spectrum with a spectral index 
of $s$ simply because the produced pions reflect the spectrum of
the relativistic protons. In the case of electrons, however, the
spectrum should be {\sl flatter} having an index $a=(s+1)/2$ --assuming that we are
away from the cutoff. This happens because the electron distribution function
should not be steepened by losses and therefore will have a slope of
index $s$ which, in turn, gives the aforementioned index in radiation.
Thus taking the usual values for $s$ between 2.1 and 2.3, one finds $a$ to vary
between 1.55 and 1.65.

Finally, it is interesting to note that, according to the present model, the $>$1 TeV flux
from SN 1006 due to primary electrons will be  $\simeq 2~10^{-12}$ ph/cm$^2$/sec
which makes this object detectable with present day Cerenkov detectors.
At the same time the low number density around SN 1006 (.05 H-atoms/cm$^3$
-- Hamilton et al. [\cite{hamiltonetal86}]) 
makes the corresponding flux due to $\pi^0$ $\gamma-$rays about an order of magnitude less.

\section{Conclusions}
Adopting a simple model for diffusive acceleration at shock fronts
we have calculated  the non-thermal luminosity of SNRs
by treating the radiation problem in a self-consistent manner;
thus we first solved
the time dependent kinetic equations for electrons
and protons and consequently we used the obtained
distribution functions to calculate the radiated spectra. 
We have shown that even for ambient matter densities as high as
1 H-atom cm$^{-3}$, the X--ray luminosity produced by the synchrotron
radiation of secondary electrons is many orders of magnitude below
the radiation of the directly accelerated (primary) electrons provided
that the latter are given about .001 of the total SN energy budget.
Thus a possible confirmation of the non-thermal character
of the X--ray flux detected from SN 1006 (Koyama et al. [\cite{koyamaetal95}]) 
will indeed indicate that the radiating electrons are directly accelerated by
the SNR blast wave and are not injected as secondaries from
hadronic collisions.

On the other hand we find that the directly accelerated electrons can
produce a TeV flux by inverse Compton scattering on the microwave
background which is about of the same order as the flux produced from
the $\pi^0$ decay (DAV94) provided that the magnetic field is not
much higher than 10 $\mu$G. 
Therefore a possible detection of a SNR at 
TeV energies will not neccessarily mean that protons are accelerated there.
To deduce that one needs detailed spectral information as electrons
are expected to give flatter power laws.

Note also that as we have assumed that the
high energy electrons scatter only on the microwave background, all the
inverse Compton fluxes presented here are only  lower limits since the 
inclusion of any other photon field 
(such as the galactic infrared background--see 
Cox \& Mezger [{\cite{coxmezger89}}])
would enhance the inverse Compton emissivity.
Interestingly enough, TeV observations of SNRs have started constraining
the theories of proton acceleration
(Lessard et al. [\cite{lessardetal95}]), so this
constraint, when applied to electron radiation alone, might be used to obtain
a limit on the magnetic field (De Jager et al. [\cite{dejageretal95}]).

Finally, one should add that our results do not simply scale
with the ambient matter density when this becomes large.
In such a case electron bremsstrahlung becomes important not only
as a radiation mechanism but as an energy loss process as well.
This effectively is an additional term
in Eqn. (\ref{ekinetic}) which will affect the electron distribution
function and thus the radiated spectrum.
\vspace{-0.3 cm}

\acknowledgements
I would like to thank Okkie de Jager, Peter Duffy and John Kirk for 
many discussions, the two referees Drs. Chevalier and Drury whose comments
helped clarify some of the issues and
the Deutsche Forschungsgemeinschaft for support under
Sonderforschungsbereich 328.

\end{document}